\documentclass[aps,prc,twocolumn,superscriptaddress,showpacs]{revtex4-1}
\usepackage{hyperref}

\usepackage[dvips]{graphicx}
\usepackage{amsmath,amssymb}
\usepackage[T1]{fontenc}

\begin{document}

\title{Inelastic neutrino scattering off hot nuclei in supernova environments}

\author{Alan~A.~Dzhioev}
\email{dzhioev@theor.jinr.ru}
\affiliation{Bogoliubov Laboratory of Theoretical Physics, JINR, 141980, Dubna, Russia}
\author{A.~I.~Vdovin}
\affiliation{Bogoliubov Laboratory of Theoretical Physics, JINR, 141980, Dubna, Russia}
 \author{J.~Wambach}
\affiliation{Institut f{\"u}r Kernphysik, Technische Universit{\"a}t Darmstadt, 64289 Darmstadt, Germany}
\affiliation{GSI Helmholtzzentrum f\"ur Schwerionenforschung, Planckstr. 1, 64291 Darmstadt, Germany}
\author{V.~Yu.~Ponomarev}
\affiliation{Institut f{\"u}r Kernphysik, Technische Universit{\"a}t  Darmstadt, 64289 Darmstadt, Germany}

\date{\today}

\begin{abstract}
We study  inelastic neutrino scattering off hot nuclei for temperatures relevant under  supernova
conditions.  The method we use is based on the quasiparticle
random phase approximation extended to finite temperatures within
the thermo field  dynamics (TQRPA).  The method allows a transparent treatment of  upward and downward transitions in hot nuclei, avoiding the application of Brink's hypothesis.
For the sample nuclei $^{56}$Fe and $^{82}$Ge we  perform a detailed analysis of thermal effects on
the strength distributions of allowed Gamow-Teller (GT) transitions which dominate the
scattering process at low neutrino energies. For  $^{56}$Fe and $^{82}$Ge the finite temperature
cross-sections are calculated by taking into account the contribution of allowed and forbidden transitions.
The observed enhancement of the cross-section at low neutrino energies is explained by considering thermal
  effects on the GT strength.
For $^{56}$Fe we compare the calculated cross-sections to those obtained earlier from
a hybrid approach that combines large-scale shell-model and RPA calculations.
\end{abstract}




\maketitle

\section{Introduction}

The significant role played by processes involving neutrinos in
core-collapse supernovae (type II supernovae) is well
known~\cite{Janka_PhysRep442}. Until the core reaches densities of
$\rho\sim 10^{11}\,\mathrm{g\,cm^{-3}}$, a substantial amount of the gravitational energy of
the collapse is radiated by neutrinos that leave the star
freely. However, at higher densities neutrino interactions with
matter become important on the time-scale of the collapse, leading to neutrino trapping and thermalization.
Supernova
core-collapse  simulations require a detailed  description of
neutrino transport and should in principle include all potentially
important neutrino reactions.

It was first pointed by Haxton~\cite{Haxton_PRL60} that the
neutral-current inelastic neutrino scattering on nuclei involving
the excitation of giant resonances can lead to significant neutrino
cross-sections and, therefore, this process should be incorporated
into core-collapse simulations. Shortly thereafter, this was done by
Bruenn and Haxton~\cite{Bruenn_APJ376}. They found that the inelastic
neutrino-scattering on nuclei plays the same important role as the
neutrino-electron scattering in equilibrating neutrinos with matter (see also Ref.~\cite{Langanke_PRL100}).

In their study Bruenn and Haxton approximated  the nuclear
composition of the core by a single representative nucleus,
$^{56}$Fe. Moreover, the relevant cross-sections were
calculated by assuming that only allowed Gamow-Teller and first-forbidden upward transitions
from the nuclear ground-state contribute to neutrino scattering. However,  supernova
matter has a temperature of an order of 1~MeV or higher
and the neutrinos scatter off nuclei which are in thermally populated excited states.
As was first realized in Ref.~\cite{Fuller_APJ376}, upward and downward transitions from nuclear excited
states to lower-lying states completely remove the energy threshold for the inelastic neutrino-nucleus
scattering in the supernova environment
and  contribute to a significant enhancement of the cross-section for low-energy neutrinos.
Moreover, and this is more important, due to downward transitions from nuclear excited states to  lower-lying states
neutrinos can gain energy after interacting with the nucleus, thereby assisting in cooling the core and reducing its entropy.
This is different to inelastic scattering with electrons where due to the degeneracy of electrons neutrinos mainly loose energy.

An  explicit calculation of reaction rates and cross-sections  at
finite temperature can be performed by summing over
Boltzmann-weighted, individually determined contributions from
nuclear excited states. However, for $T\gtrsim1\,\mathrm{MeV}$ a
state-by-state evaluation includes too many states to derive the cross
section for each individual state and, hence, is unfeasible. To
overcome this difficulty an approximate method to treat thermal
effects on the inelastic neutrino-nucleus scattering was proposed
in~\cite{Juodagalvis_NPA747} (see also Ref.~\cite{Sampaio_PLetB529})
within the so-called hybrid approach~\cite{Kolbe_PRC63, Toivanen_NPA694}.
In this method the contributions of the allowed Gamow-Teller
transitions to the neutrino-nucleus cross-section are derived from large-scale shell-model (SM)
calculations, while the forbidden contributions are considered within the RPA.

To treat thermal effects within the hybrid approach,
the Gamow-Teller contribution to the cross-section is split into
the neutrino down-scattering ($E_{\nu'}<E_\nu$) and
neutrino up-scattering ($E_{\nu'}>E_\nu$) parts,
where $E_\nu,~E_{\nu'}$ denote the neutrino energies in the initial and final states, respectively.
For the down-scattering part the Brink hypothesis was applied which
states that  GT distributions built on nuclear excited states are
the same as those for the  nuclear ground state but shifted by the
excitation energy. Under this assumption, the down-scattering part
of the cross-section becomes temperature independent. The
temperature dependence arises from the up-scattering part which
accounts for contributions of downward transitions from nuclear excited
states. These contributions are determined  by 'inversion' of the
shell-model GT distributions for the low-lying states.

Large-scale shell-model  calculations provide a detailed strength
distribution of charge-neutral Gamow-Teller (GT$_0$) transitions that strongly dominate the inelastic
neutrino-nucleus scattering at low neutrino energies ($E_\nu\lesssim
15\,\mathrm{MeV}$). However, being applied  to hot nuclei,  this method has
its own shortcomings mainly because it partially employs the Brink
hypothesis when treating GT$_0$ transitions from nuclear excited states.
As follows from experimental studies of giant dipole resonances
(GDR) in hot nuclei, the GDR  strength function exhibits a
temperature dependence (see, e.g., the monograph \cite{Bortignon1998}
and one of the latest reviews~\cite{Santonocita_EPJA30}), i.e., the
validity of the Brink hypothesis is not obvious. Moreover,
theoretical calculations performed for charge-exchange GT
transitions in the framework of the shell-model Monte Carlo (SMMC)
method demonstrate that with increasing temperature the GT centroid
shifts to lower energies and the width of the
distribution increases with the appearance of low-lying states~\cite{Radha_PRC56}.
In addition, the present computer capabilities allow application of
large-scale shell-model calculations only to iron group
nuclei ($pf$-shell, $A=45-65$), whereas  neutrino scattering on
more massive and neutron-rich nuclei also may play an important role
in various astrophysical scenarios. Thus, the problem of an accurate
description of inelastic neutrino-nucleus scattering in the supernova
environment is not solved completely yet and alternative methods are
desirable.

In~\cite{Dzhioev_PAN74}, we have developed such an alternative approach to
treat thermal effects on inelastic  neutrino-nucleus scattering
cross-sections. This approach is based on the thermal quasiparticle
random phase approximation (TQRPA). We apply it in the context of
the thermo field dynamics (TFD) formalism~\cite{Takahashi_IJMPB10,Umezawa1982,Ojima_AnPhys137}
which enables a transparent treatment of upward  and downward transitions
from thermally excited nuclear states and opens possibilities for
systematic improvements. This approach was also recently used in
studies of the electron capture on hot nuclei under supernova
conditions~\cite{Dzhioev_PRC81}.

In~\cite{Dzhioev_PAN74}, the thermal effects on the inelastic neutrino
scattering off the hot $^{54}$Fe nucleus were investigated. It was
shown that the TQRPA does not support Brink's hypothesis and leads
to temperature dependent strength distributions for allowed and
forbidden transitions. As a result, both the up- and down-scattering
parts of the cross-section are temperature dependent. Despite the
differences between the two approaches, the TQRPA revealed the same
thermal effect as was found in~\cite{Juodagalvis_NPA747}. Namely, a
temperature increase results in considerable enhancement of the
cross-section for neutrino energies lower than the energy of the
GT$_0$ resonance.

In the present paper, we extend our previous study by also considering inelastic
neutrino scattering off neutron-rich nuclei beyond $pf$-shell. In our calculations,
we take into account not only the
first-forbidden transitions but also contributions from higher
multipoles. For the selected iron isotope, ${}^{56}$Fe, we perform a
detailed comparison of the calculated TQRPA 
cross-sections with the hybrid approach results and discuss the reason
for the observed discrepancy  at low neutrino energies.

The paper is organized as follows. In Sec.~\ref{formalism}  we
present some important features of the TFD formalism and briefly
outline how to treat upward and downward transitions in a hot nucleus
within the TQRPA. The details of our approach are expounded
in~\cite{Dzhioev_PRC81,Dzhioev_PAN74,Dzhioev_IJMPE18}. In
Sec.~\ref{formalism}, we also provide the necessary formulas to calculate
inelastic neutrino-nucleus cross-sections at finite temperatures.
The results of the numerical calculations are presented and discussed in
Sec.~\ref{results} for the sample nuclei $^{56}$Fe and $^{82}$Ge.
The conclusions are summarized in Sec.~\ref{conclusion}.

\section{Formalism}\label{formalism}

In the stellar environment during the core-collapse phase all
nuclear reactions mediated by the strong and electromagnetic
interaction are in equilibrium with their
inverse~\cite{Janka_PhysRep442}. Neglecting weak-interaction
mediated reactions, nuclei are in thermal equilibrium with heat and
particle reservoirs and, therefore, can be described as a thermal
ensemble. In TFD, such an equilibrium ensemble is represented by
a temperature-dependent state termed the thermal vacuum
$|0(T)\rangle$ \footnote{The correspondence between the thermo field
dynamics and the superoperator formalism is discussed
in~\cite{Schmutz_ZPhysB30}. The latter is used by one of the authors (A.D.) to
study nonequilibrium transport phenomena (see,
e.g.,~\cite{Dzhioev_JPhys24})}.
 The thermal vacuum is  determined as the zero-energy eigenstate of the thermal Hamiltonian,
${\mathcal H} = H - \widetilde H$, and it satisfies the thermal state condition
\begin{equation}\label{TSC}
  A|0(T)\rangle = \sigma_A \mathrm{e}^{\mathcal{H}/2T}\widetilde A^\dag |0(T)\rangle.
\end{equation}
In the above equations $H$ is the original nuclear Hamiltonian  and
$\widetilde H$ is its tilde counterpart acting in the auxiliary
Hilbert space; an operator $A$ acts in the physical Hilbert space,
$\widetilde A$ is its tilde partner, and $\sigma_A$ is a phase factor.
The thermal state condition guarantees that the expectation value
$\langle 0(T)|A|0(T)\rangle$ is equal to the (grand)canonical
average of $A$. In this sense,  relation~\eqref{TSC} is
equivalent to the Kubo-Martin-Schwinger condition for an equilibrium
(grand)canonical density matrix~\cite{Kubo_JPSJ12,*Martin_PRev115}.

Weak-interaction processes, such as inelastic neutrino scattering,
induce transitions from the thermal vacuum to excited states of
the thermal nuclear Hamiltonian. As follows from the definition
of ${\mathcal H}$, each of its eigenstates with positive energy has
a counterpart --- the tilde-conjugate eigenstate --- with negative
but same absolute value of energy. Transitions from the thermal
vacuum to positive energy states (upward transitions) correspond to
excitation of the nucleus, while transitions to negative energy
states (downward transitions) describe the decay of thermally excited
states.

\subsection{Thermal quasiparticle RPA}\label{TQRPA}

Let us now consider a general nuclear Hamiltonian consisting of
mean fields for protons and neutrons, pairing interactions, and
residual two-nucleon interactions:
\begin{equation}
  H = H_\mathrm{mf} +  H_\mathrm{pair} + H_\mathrm{res}.
\end{equation}
To fix an average number of protons and neutrons we introduce the
respective chemical potentials into~$H_\mathrm{mf}$. The
residual interaction can contain  both particle-hole and
particle-particle terms. We  assume a spherically symmetric nucleus,
although the deformation can be easily included into the theory.
Within the TQRPA,  to find excited states of a hot nucleus, we first
introduce thermal quasiparticle creation
($\beta^\dag,~\widetilde\beta^\dag$) and annihilation
($\beta,~\widetilde\beta$) operators which account for pairing correlations at finite
temperature. The structure of this operators is found by
diagonalizing the
$\mathcal{H}_\mathrm{mf} + \mathcal{H}_\mathrm{pair}$ part of the
thermal Hamiltonian and simultaneously demanding that the vacuum of thermal quasiparticles obeys
the thermal state condition~\eqref{TSC}. Then, to account for the long-range residual
interaction, we introduce  thermal phonon operators
$Q^\dag_{JMi},~\widetilde Q^\dag_{JMi}$ of given total
angular momentum $(J,M)$ whose action on the thermal
vacuum $|0(T)\rangle$ creates thermal excited states, while the
thermal vacuum itself is the vacuum for the $Q_{JMi},~\widetilde
Q_{JMi}$ operators.

The structure and the energy of thermal phonons can be found by
applying either the variational principle or the equation of motion
method under two additional constraints: (i)  phonon operators
commute like bosonic ones; (ii) the vacuum of thermal phonons obeys
the thermal  state condition~\eqref{TSC}. The resulting phonon
operators have the following form \footnote{In Eq.~\eqref{phonon},
$[\,]^J_M$ denotes the coupling of two single-particle angular
momenta $j_1,\,j_2$ to the total angular~momentum~$J$. The bar over
index $j$ means time-inversion.}:
\begin{multline}\label{phonon}
  Q^\dag_{J M i}=\sum_{j_1j_2}
 \Bigl\{\psi^{Ji}_{j_1j_2}[\beta^\dag_{j_1}\beta^\dag_{j_2}]^J_M +
 \widetilde\psi^{J i}_{j_1j_2}[\widetilde\beta^\dag_{\overline{\jmath_1}}
 \widetilde\beta^\dag_{\overline{\jmath_2}}]^J_M
 \\ +
 \eta^{J i}_{j_1j_2}[\beta^\dag_{j_1}
  \widetilde\beta^\dag_{\overline{\jmath_2}}]^J_M
+
 \phi^{J i}_{j_1j_2}[\beta_{\overline{\jmath_1}}\beta_{\overline{\jmath_2}}]^J_M  \\
+ \widetilde\phi^{J i}_{j_1j_2}[\widetilde\beta_{j_1}
 \widetilde\beta_{j_2}]^J_M +
  \xi^{J i}_{j_1j_2}[\beta_{\overline{\jmath_1}}
  \widetilde\beta_{j_2}]^J_M\Bigr\}
\end{multline}
and they diagonalize the thermal  Hamiltonian
\begin{equation}
{\cal H}\simeq\sum_{JM i}\omega_{J i}(T)
   (Q^\dag_{JM i}Q^{\phantom{\dag}}_{JM i}
   -\widetilde Q^\dag_{JM i}\widetilde Q^{\phantom{\dag}}_{JM i})
\end{equation}
 within the TQRPA. The phonon amplitudes $\psi,\, \widetilde\psi,\,\mathrm{etc.}$
 as well as the phonon energies $\omega$ are
 the solution of the TQRPA equations. It should be emphasized
 that in the zero-temperature limit the TQRPA method turns into
 the standard QRPA.

In~\cite{Dzhioev_PRC81}, we have performed a detailed analysis of
finite temperature effects on the spectrum of charge-exchange
thermal phonons. Here we repeat the main conclusions which remain
valid for the charge-neutral excitations as well. Due to the terms
in~\eqref{phonon}  involving  tilde thermal quasiparticle operators
(terms like $\beta^\dag\widetilde\beta^\dag$ and
$\widetilde\beta^\dag\widetilde\beta^\dag$)  the spectrum of thermal
phonons contains negative- and low-energy states which do not exist
at zero temperature.  Since (see~\cite{Dzhioev_PRC81} for more
details) the creation of a tilde quasiparticle is equivalent  to the
annihilation of a thermally excited Bogoliubov quasiparticle, the
excitation of the aforementioned "new" phonon states can be interpreted
as thermally unblocked transitions from nuclear excited states.
Furthermore, both the energies of thermal quasiparticles and the
interaction strength between them are temperature dependent. As a
result, after solving the TQRPA equations we obtain a temperature-dependent spectrum
of thermal phonons.

Once the structure of thermal phonons is determined,  one can
evaluate transition strengths (probabilities) from the thermal
vacuum to thermal one-phonon states. For a given transition operator
$\mathcal{T}$ we have
\begin{align}\label{trans_ampl}
\Phi_{J i}&=\bigl|\langle Q_{J
i}\|\mathcal{T}\|0(T)\rangle\bigr|^2,
  \notag\\
\widetilde \Phi_{J i}&=\bigl|\langle\widetilde Q_{J
i}\|\mathcal{T}\|0(T)\rangle\bigr|^2,
\end{align}
where $\Phi_{J i}$ and $\widetilde \Phi_{J i}$ are the strengths of
upward and downward transitions, respectively. They are connected by
the relationship
\begin{equation}\label{balance}
\widetilde\Phi_{J i}=\exp\Bigl(-\frac{\omega_{J i}}{T}\Bigr)\Phi_{J
i},
\end{equation}
where $\omega_{Ji}$ is a positive solution of the TQRPA equations.
This relation links the probabilities to transfer and gain energy $E=\omega_{Ji}$ from
a hot nucleus. It is interesting to note that the same relationship between the upward and
downward transition strengths is used in~\cite{Fischer_arxiv1309.4271v1} when considering
the thermal strength functions for emission and absorption of
neutrino-antineutrino pairs by hot nuclei.
In~\cite{Fischer_arxiv1309.4271v1}, the relation results from the principle of
detailed balance. In TFD, it arises from  the thermal state
condition imposed on the thermal vacuum.
We also would like to point out that in~\cite{Fischer_arxiv1309.4271v1}, due to the application of Brink's hypothesis, the absorption (upward) strength
is considered to be temperature independent and only the emission (downward) strength depends on temperature due to the Boltzmann factor~$\exp(-E/T)$.
In contrast, within the present approach, both the upward and downward transition strengths are temperature dependent.

\subsection{Cross-section at finite temperatures}

Deriving the inelastic neutrino-nucleus cross-section at finite
temperature we follow the Walecka-Donnelly
formalism~\cite{Walecka1975,Donnelly_PRep50} which is based on the standard
current-current form of the weak interaction Hamiltonian. Then the
temperature-dependent differential cross-section for a transition from the thermal vacuum
to the final thermal one-phonon state takes the form
\begin{equation}\label{dif_cr_sect}
  \frac{d}{d\Omega} \sigma_{Ji}(E_\nu,T)= \frac{2G^2_F}{\pi}\, E_{\nu'}^2 \cos^2\frac{\Theta}{2}
    \bigl\{\sigma^J_{CL} + \sigma^J_{T}   \bigr\}.
\end{equation}
Here, $G_F$ is the Fermi constant of the weak interaction and  $\Theta$
the scattering angle. The Coulomb-longitudinal,
$\sigma^J_{CL}$, and transverse, $\sigma^J_{T}$, terms in
Eq.~\eqref{dif_cr_sect} are given by
\begin{equation}\label{CL}
  \sigma^J_{CL} = |\langle J i\| \hat M_J \pm \frac{\omega_{Ji}}{q} \hat L_J\|0(T)\rangle|^2,
\end{equation}
\begin{multline}\label{T}
 \sigma^J_{T}=\Bigl(-\frac{q^2_\mu}{2q^2} + \tan^2\frac{\Theta}{2} \Bigr)
  \Bigl[ |\langle J i\| \hat T^\mathrm{mag}_J\|0(T)\rangle|^2 +\\
   |\langle J i\| \hat T^\mathrm{el}_J\|0(T)\rangle|^2       \Bigr]
  -\tan\frac{\Theta}{2}\sqrt{-\frac{q^2_\mu}{2q^2} + \tan^2\frac{\Theta}{2}} \\
    \times\Bigl[ 2 \mathrm{Re} \langle J i\| \hat T^\mathrm{mag}_J\|0(T)\rangle\langle J i\| \hat T^\mathrm{el}_J\|0(T)\rangle^*  \Bigr],
\end{multline}
where $q_\mu=(\pm\omega_{J i}, \vec q)$ ($q=|\vec q|=\sqrt{\omega^2_{Ji}+4E_{\nu'}E_\nu\sin^2\frac{\Theta}{2})}$) is the 4-momentum
transfer and the notation $|Ji\rangle$ is used to denote both the
non-tilde and the tilde states. The upper sign in the above equations refers to upward transitions from the thermal
vacuum to non-tilde states ($E_{\nu'}=E_\nu - \omega_{Ji}$), while the lower sign corresponds to
downward transitions to tilde states ($E_{\nu'}=E_\nu + \omega_{Ji}$).
The multipole operators $\hat M_J$, $\hat L_J$,  $\hat J^\mathrm{el}_J$, and $\hat J^\mathrm{mag}_J$ denote
the charge, longitudinal, and transverse electric and  magnetic
parts of the hadronic current, respectively, as defined
in~\cite{Walecka1975,Donnelly_PRep50}. For the vector, axial-vector,  and pseudoscalar form-factors which describe
the internal structure of the nucleon we use parametrization from Ref.~\cite{Singh_NPB77} (see also Ref.~~\cite{Djapo_PRevC86}).

From Eq.\,\eqref{dif_cr_sect}, the total cross-section,
$\sigma(E_\nu,T)$, as a function of temperature and incoming
neutrino energy is obtained by integrating over the scattering angle and
summing over all possible final thermal excited states
\begin{align}\label{total_CrSect}
  \sigma(E_\nu,T) = &~ 2\pi\sum_{Ji}\int^{-1}_{1} \frac{d\sigma_{Ji}}{d\Omega}\, d\cos\Theta
  \notag\\
  &~ = \sigma_\mathrm{down}(E_\nu,T) + \sigma_\mathrm{up}(E_\nu,T).
\end{align}
Here, we follow Ref.~\cite{Juodagalvis_NPA747} and split the total
cross-section  into  two parts: $\sigma_\mathrm{down}(E_\nu,T)$
describes the neutrino down-scattering process and includes only
transitions to non-tilde phonon states, while
$\sigma_\mathrm{up}(E_\nu,T)$ corresponds to the neutrino up-scattering
associated with transitions to tilde states.

For inelastic scattering of low-energy neutrinos, i.e., in the long wavelength limit ( $q\to 0$), only two multipole operators survive,
$\hat L_1$ and $\hat T^\mathrm{el}_1$, which contribute to $1^+$ transitions. Then the integration over the scattering angle in Eq.~\eqref{total_CrSect}
can be performed analytically and, in view of the detailed balance principle~\eqref{balance}, the low-energy cross-section can be written as
\begin{align}\label{cr_sect_GT0}
  \sigma(E_\nu,T) =&\frac{G^2_F}{\pi}{\sum_i}' (E_\nu - \omega_{i})^2\Phi_{i}
  \notag\\
  +&  \frac{G^2_F}{\pi}\sum_i (E_\nu + \omega_{i})^2\exp\Bigl(-\frac{\omega_{i}}{T}\Bigr)\Phi_{i},
\end{align}
where $\Phi_{i}$  is the transition strength for the Gamow-Teller operator  (see
Eq.~\eqref{GT0}  below). The  sum ${\sum_i}'$ in the first, down-scattering, term  implies
summation over  $1^+$ non-tilde thermal phonon states with the energy $\omega_{i}<E_\nu$.
Apparently, for vanishing neutrino energies,  $E_\nu\approx 0$, only the second,  up-scattering, term persists at finite temperatures. We also note that although
the Boltzmann factor suppresses the contributions of downward transitions from high-lying thermally excited states, the phase-space factor $(E_\nu + \omega_{i})^2$ acts in the opposite
direction and favors them.

\section{Results  and discussion}\label{results}

The formalism presented above is employed to study thermal effects
on the inelastic neutrino scattering off the two sample nuclei,
$^{56}\mathrm{Fe}$ and $^{82}\mathrm{Ge}$. The iron isotope is among
the most abundant nuclei at the early stage of the core-collapse,
while the neutron-rich  germanium isotope can be considered
as the average nucleus at later stages~\cite{Cooperstein_NPA420}.

Let us now specify  the nuclear Hamiltonian which will be used in
the present study. Like in~\cite{Dzhioev_PAN74,Dzhioev_PRC81}, we
apply a phenomenological Hamiltonian containing separable particle-hole residual interactions with
isoscalar and isovector parts. We neglect  particle-particle
interactions except for the BCS pairing forces. This Hamiltonian is usually
referred to as the quasiparticle-phonon model (QPM)~\cite{Soloviev1992}.
For $^{56}\mathrm{Fe}$ and $^{82}\mathrm{Ge}$, the single-particle
energies and wave-functions are derived from an appropriate
Woods-Saxon mean-field potential~\cite{Chepurnov_SJNP6}. The depth of the
Wood-Saxon potential as well as the pairing strength parameters are
fixed in the same manner as in~\cite{Dzhioev_PAN74,Dzhioev_PRC81}.
In the obtained single-particle level schemes $^{56}\mathrm{Fe}$ has
two neutron holes in the $1f_{7/2}$ subshell and two protons in the
$2p_{3/2}$ subshell, while $^{82}\mathrm{Ge}$ has closed $1g_{9/2}$
neutron and $2p_{3/2}$ proton subshells. It is notable that the
sequence of single-particle levels  for $^{82}\mathrm{Ge}$ is close
to that used in Ref.~\cite{Cooperstein_NPA420} for the same nucleus in spite of
different Woods-Saxon parameterizations. Solving
the BCS equations at zero-temperature we get the proton and neutron
pairing gaps:
 $\Delta_{p(n)} = 1.57(1.36)~\mathrm{MeV}$ for $^{56}\mathrm{Fe}$
and $\Delta_{p(n)} = 1.22(0.0)~\mathrm{MeV}$ for
$^{82}\mathrm{Ge}$.
Thus, the critical temperature $T_\mathrm{cr}\approx 0.5\Delta(T=0)$
above which the pairing gap collapses,  according to the BCS theory
(see Refs.~\cite{Goodman_NPA352,Civitarese_NPA404} for more details),
 is $T_\mathrm{cr}\approx 0.8\,\mathrm{MeV}$ for $^{56}$Fe and
 $T_\mathrm{cr}\approx 0.6\,\mathrm{MeV}$ for $^{82}$Ge.

In the present study, multipoles up to $J^\pi = 3^\pm$
contributing to the neutrino-scattering cross-section~\eqref{dif_cr_sect} are included
in the calculations. To generate the thermal one-phonon excited states, we use both
multipole and spin-multipole
components of the residual interaction
\begin{align}\label{mult}
H^{\rm ph}_{\rm M}&=-\frac12\sum_{\lambda\mu}\sum_{\genfrac{}{}{0pt}{1}{\tau=n,p}{\rho=\pm1}}
 (\kappa_0^{(\lambda)}+\rho\kappa_1^{(\lambda)})M^\dag_{\lambda\mu}(\tau)M^{\phantom{+}}_{\lambda\mu}(\rho\tau),
 \notag   \\
 H^{\rm ph}_{\rm SM} &=
-\frac12\sum_{L\lambda\mu}\sum_{\genfrac{}{}{0pt}{1}{\tau=n,p}{\rho=\pm1}}(\kappa^{(L\lambda)}_0+\rho\kappa^{(L\lambda)}_1)
S^\dag_{L\lambda\mu}(\tau)
S^{\phantom{\dag}}_{L\lambda\mu}(\rho\tau).
\end{align}
Here $M^\dag_{\lambda\mu}$ and $S^\dag_{L\lambda\mu}$ are
single-particle multipole and spin-multipole operators~\cite{Soloviev1992},
and changing the sign of the isotopic index $\tau$ means changing $n
\leftrightarrow p$. The excitations of natural parity ($\pi=(-1)^J$)
are generated by the multipole and spin-multipole $L=\lambda$
interactions~\eqref{mult}, while the spin-multipole interactions
with $L=\lambda\pm1$ are responsible for the states of unnatural
parity ($\pi=(-1)^{J+1}$). To generate $0^+$ excitations, we take into account both the particle-hole residual interaction and the particle-particle interaction stemming from the pairing part of the Hamiltonian.
Here we would like to emphasize that the inclusion of the particle-particle
residual interaction into the Hamiltonian does not affect the strength distributions and the cross-sections
 for temperatures above the critical one.

In contrast to~\cite{Dzhioev_PAN74,Dzhioev_JPhys410}, in the present study the radial
form-factors of multipole and spin-multipole operators in
Eq.~\eqref{mult} have the $r^\lambda$ form. We  found that this form of the radial form-factors gives
better agreement with results of relativistic  self-consistent QRPA calculations~\cite{Djapo_PRevC86} when
comparing multipole composition of the cross-sections (see the discussion below). The respective isoscalar
and isovector strength parameters, $\kappa_{0,1}^{(\lambda)}$ and
$\kappa_{0,1}^{(L\lambda)}$, are first roughly estimated following
Refs.~\cite{Castel_PLetB65,Bes_PRep16} and then partly refined on the basis of
available experimental data. For example, in $^{56}$Fe the isovector
strength parameters $\kappa^{(01)}_1$ and $\kappa^{(21)}_1$ are
slightly readjusted to reproduce the experimental centroid energies of
the GT$_-$ and GT$_+$
resonances~\cite{Rapaport_NPA410,Ronnqvist_NPA563}. We find that
the isovector strength parameter $\kappa^{(1)}_1$ estimated
according to~\cite{Bes_PRep16} reproduces the experimental
position of the GDR centroid ($\sim18~\mathrm{MeV}$) in
$^{56}$Fe~\cite{Bowles_PRevC24} quite well. In addition, the isoscalar strength
parameters $\kappa_0^{(1)}$ for $^{56}$Fe and $^{82}$Ge are fitted
to exclude the spurious $1^-$ state due to the center of mass motion
of the nucleus.

\subsection{Zero temperature}

\begin{figure}[t!]
 \begin{centering}
\includegraphics[width=\columnwidth, angle=0]{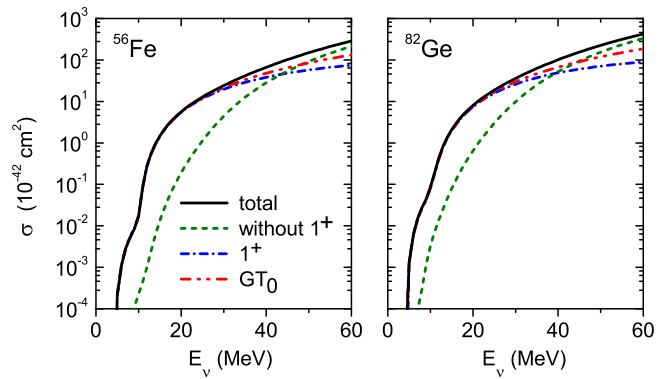}
\caption{(Color online) Inelastic neutrino scattering cross-sections off the ground
states of $^{56}$Fe and $^{82}$Ge as functions of the incoming
neutrino energy $E_\nu$. The total cross-sections include
contributions of $J^\pi = 0^\pm - 3^\pm$ multipoles (solid lines).
The dashed lines show the cross-sections calculated when the $1^+$
contributions are omitted. The dash-dotted lines display the $1^+$
contributions to the cross-sections calculated with the full
$q$-dependent transition operator whereas the $1^+$ contributions
calculated with the GT$_0$ operator~\eqref{GT0} are shown by the
dash-double dotted lines.}
\label{CrSect_T0}
 \end{centering}
\end{figure}

Before proceeding to thermal effects we consider the inelastic
neutrino-nucleus cross-sections at zero temperature  and perform a
comparison with the available results of other approaches. We note once again
that at $T=0$ our calculations are equivalent to the QRPA. The
calculated ground-state cross-sections for $^{56}$Fe and $^{82}$Ge
are shown in Fig.~\ref{CrSect_T0} for incoming neutrino energies
$E_\nu=0-60\,\mathrm{MeV}$. As one can see in the figure, for
neutrinos with $E_\nu<30\,\mathrm{MeV}$ the total cross-sections are
dominated by $1^+$ transitions. Due to the energy gap in the $1^+$
nuclear states the cross-sections drop rapidly to zero as the
neutrino energy approaches the reaction threshold. Within the
present QRPA calculations, the lowest $1^+$ states in $^{56}$Fe and
$^{82}$Ge have energies 4.06 and 2.67~MeV, respectively. Note, that
the experimental energy of the first $1^+$ excited state in
$^{56}$Fe is $3.12$~MeV.

For the ground-state cross-sections we also analyze the effect due
to the exploitation of  the full $q$-dependent $1^+$ transition operator
instead of its long wavelength limit. In the latter case
the $1^+$ operator reduces to the Gamow-Teller operator
\begin{equation}\label{GT0}
  \mathrm{GT}_0 =\Bigl(\frac{g_A}{g_V}\Bigr)\vec\sigma t_0,
\end{equation}
where $(g_A/g_V) =-1.2599$~\cite{Towner1995}  is the ratio of the axial
and vector weak coupling constants and $t_0$ denotes the zero
component of the isospin operator in spherical coordinates.
Here we would like to remind that within the hybrid approach~\cite{Kolbe_PRC63, Toivanen_NPA694, Juodagalvis_NPA747},
the GT contribution to the cross-section is obtained by using the large-scale shell-model calculations.
Therefore, to make
a comparison with the hybrid approach calculation more transparent, we use the same quenching
factor for the axial weak coupling constant, $g^*_A=0.74g_A$, when
calculating the matrix elements of the $1^+$ transition operator.

\begin{figure}[t!]
 \begin{centering}
\includegraphics[width=\columnwidth, angle=0]{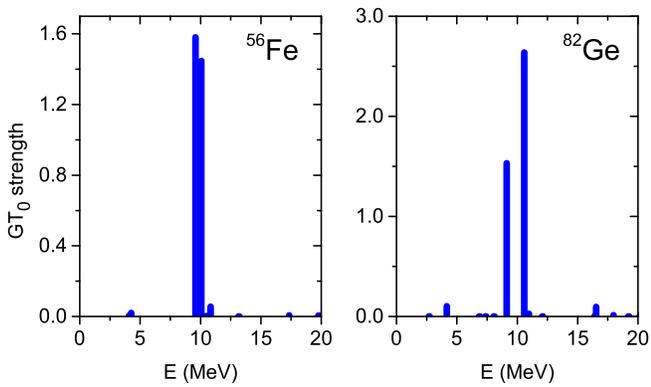}
\caption{(Color online) The distributions of the GT$_0$ strength in $^{56}$Fe and $^{82}$Ge.}
\label{GT0_0}
 \end{centering}
\end{figure}

Let us first demonstrate  the calculated QRPA (quenched)
GT$_0$ strength distributions. Referring to Fig.~\ref{GT0_0},  at zero temperature
the GT$_0$ strength is  concentrated in the resonance state around
$E=10\,\mathrm{MeV}$. According to our QRPA calculations, the main
contribution to the GT$_0$ resonance in $^{56}$Fe comes from the
proton and neutron single-particle transitions $1f_{7/2}\to
1f_{5/2}$. In $^{82}$Ge, the neutron transition
$1g_{9/2}\to1g_{7/2}$ also contributes to the resonance. In addition, for  both nuclei
our QRPA calculations predict  a weak low-lying GT$_0$ strength
($E\approx 4\,\mathrm{MeV}$) arising from the $2p_{3/2}\to 2p_{1/2}$
single-particle transitions. For $^{56}$Fe, the gross-structure of the GT$_0$
strength distribution agrees quite well with the shell-model
results~\cite{Toivanen_NPA694}, meaning that the GT$_0$ strength is
concentrated in the resonance region around 10~MeV with a small bump
at low energy. The same good agreement can be found if we compare the shell-model GT$_0$
distributions for $^{54}$Fe (see Ref.~\cite{Toivanen_NPA694}) with our previously reported QRPA result~\cite{Dzhioev_PAN74}.
However, although our calculations reproduce the resonance positions in $^{54,56}$Fe,
it is a well-known fact that the QRPA fails to recover all nuclear correlations needed
to correctly describe the full resonance width and produces only a part of it, the
so-called Landau width. The latter is quite small for the GT$_0$ resonance. As a result,  the fine structure
of GT distributions in the vicinity of the resonance is not reproduced in our calculations. In this respect the shell-model calculations are clearly advantageous.

Using the calculated strength distributions we apply Eq.~~\eqref{cr_sect_GT0} and calculate
the GT contribution to the ground-state cross-sections. In  Fig.~\ref{CrSect_T0}, these contributions are
shown by the  dash-double dotted lines.  From the figure we conclude that
for neutrino energies $E_\nu<20$~MeV, when  $1^+$ transitions
dominate the cross-section, application of the GT$_0$ operator instead of
the $q$-dependent $1^+$ operator is fully justified. However, for
neutrinos with the energy $30~\mathrm{MeV}< E_\nu< 60~\mathrm{MeV}$
the GT$_0$ operator overestimates the cross-sections by about 25\%.

\begin{table}[t!]
    \caption{The cross-sections (in units of $10^{-42}~\mathrm{cm}^2$)
    for inelastic neutrino scattering on the ground state of $^{56}$Fe.
    The present QRPA results (second column) are compared with
    those from~\cite{Djapo_PRevC86,Chasioti_PPNP59}
     and with the hybrid approach results~\cite{Kolbe_PRC63}. }
  \begin{tabular}{ccccc}
    \hline\hline
    $E_\nu$ (MeV) & QRPA & QRPA\cite{Djapo_PRevC86} & QRPA\cite{Chasioti_PPNP59} & Hybrid\cite{Kolbe_PRC63} \\
    \hline\hline
    10 & 1.69(-2)  & 1.87(-1) & 1.01(+0) & 1.91(-1)   \\
    20 & 5.64(+0)  & 9.78(+0) & 5.79(+0) & 6.90(+0)   \\
    30 & 2.41(+1)  & 4.08(+0) & 1.87(+1) & 2.85(+1)   \\
    40 & 6.65(+1)  & 1.05(+2) & 5.51(+1) & 7.86(+1)   \\
    50 & 1.49(+2)  & 2.16(+2) & 1.43(+2) & 1.72(+2)   \\
    60 & 2.87(+2)  & 3.89(+2) & 3.09(+2) & 3.20(+2)   \\
    70 & 4.83(+2)  & 6.33(+2) & 5.63(+2) & 5.25(+2)   \\
    80 & 7.36(+2)  & 9.59(+2) & 8.82(+2) & 7.89(+2)   \\
    90 & 1.03(+3)  & 1.38(+3) & 1.22(+3) & 1.11(+3)   \\
   100 & 1.36(+3)  & 1.92(+3) & 1.52(+3) & 1.49(+3)   \\
   \hline\hline
  \end{tabular}
 \label{56Fe_comparison}
\end{table}

In Table~\ref{56Fe_comparison}, we compare the calculated
ground-state  cross-sections  for $^{56}$Fe with those obtained with
the hybrid approach~\cite{Kolbe_PRC63}, the relativistic self-consistent
QRPA~\cite{Djapo_PRevC86}, and the QRPA-based framework
from~Ref.~\cite{Chasioti_PPNP59}.
The range of incoming neutrino
energies is $10\le E_\nu\le 100$~MeV.
 As it follows from the table, except for low neutrino energies ($E_\nu=10$~MeV),
 the cross-sections of all four models are in good qualitative agreement.
 It is interesting to note that for  $E_\nu\ge20$~MeV the present results are
 generally closer to the hybrid approach results than the results of the other two QRPA-based methods.

To explain the discrepancy between our calculations and those of the hybrid approach
 at low neutrino energies, we note that at $E_\nu\approx 10\,\mathrm{MeV}$
the calculated cross-sections are strongly sensitive to the
fine details of the GT$_0$ distribution in the resonance region. As
it was already discussed above, the large-scale shell-model calculations adequately
reproduce the fragmentation of the GT$_0$ resonance strength
whereas the QRPA calculations predict its much stronger
concentration near the excitation energy $E\approx
10\,\mathrm{MeV}$. For this reason, our cross-section calculated for
$E_\nu= 10\,\mathrm{MeV}$ is considerably smaller than the
hybrid approach result.

\begin{figure}[b!]
 \begin{centering}
\includegraphics[width=\columnwidth, angle=0]{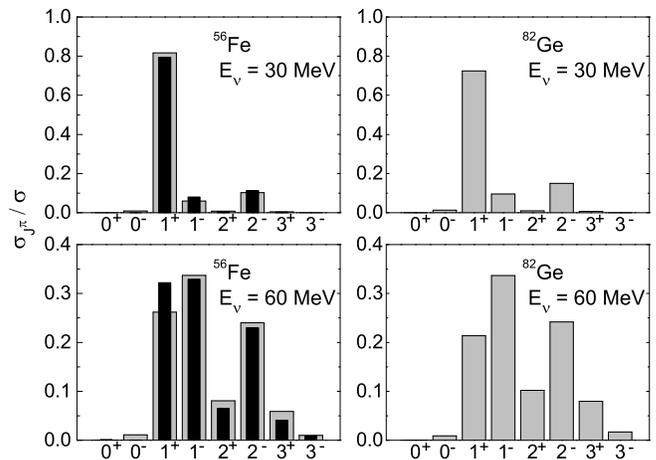}
\caption{Relative contributions of multipole transitions $J^\pi =
0^\pm-3^\pm$ in the cross-sections for the inelastic neutrino
 scattering on the ground-states of $^{56}$Fe and $^{82}$Ge at
 incoming neutrino energies $E_\nu=30$ and 60~MeV. For $^{56}$Fe, the results
 of the present analysis (grey bars) are compared with the results of relativistic self-consistent QRPA calculations (black bars). The latter are
 obtained from Fig.~3 in Ref.~\cite{Djapo_PRevC86}.}
\label{Mult_contributions}
 \end{centering}
\end{figure}

We also study how relative contributions
of different multipoles to the total cross-sections depend on the energy of incoming
neutrinos. In Fig.~\ref{Mult_contributions},  the relative contributions are shown for
$E_\nu=30~\mathrm{and}~60~\mathrm{MeV}$ neutrinos. Even at
$E_\nu=30~\mathrm{MeV}$ a largely dominant multipole is $1^+$,
although  contributions coming from the other multipoles are not
negligible. For $^{82}$Ge this contribution reaches about 30\% of
the total cross-section. This is due to the neutron excess which makes
possible spin-dipole $1^-$ and $2^-$ $1\hbar\omega$ transition at
relatively low neutrino energies. The situation is quite different
for $E_\nu=60~\mathrm{MeV}$ where the multipole transitions $J^\pi = 1^+,\,1^-\,\mathrm{and}\,2^-$ contribute
about equally to the cross-sections.

In Fig.~\ref{Mult_contributions}, we compare the obtained
multipole composition of the cross-section for $^{56}$Fe with that from relativistic self-consistent QRPA calculations~\cite{Djapo_PRevC86}.
Although our cross-sections are
somewhat smaller than those in Ref.~\cite{Djapo_PRevC86} (compare the second and the third columns of Table~\ref{56Fe_comparison}), one can observe an
excellent agreement between the two
models based on rather different backgrounds.  In accordance with Ref.~\cite{Djapo_PRevC86},
we find that $0^+$  allowed transitions only marginally contribute
to the inelastic cross-section and this finding is true  for finite
temperatures as well. For this reason, in the discussion below, we
will always mean only the $1^+$ multipole channel when considering the
allowed transitions.

\begin{figure}[t!]
 \begin{centering}
\includegraphics[width=\columnwidth, angle=0]{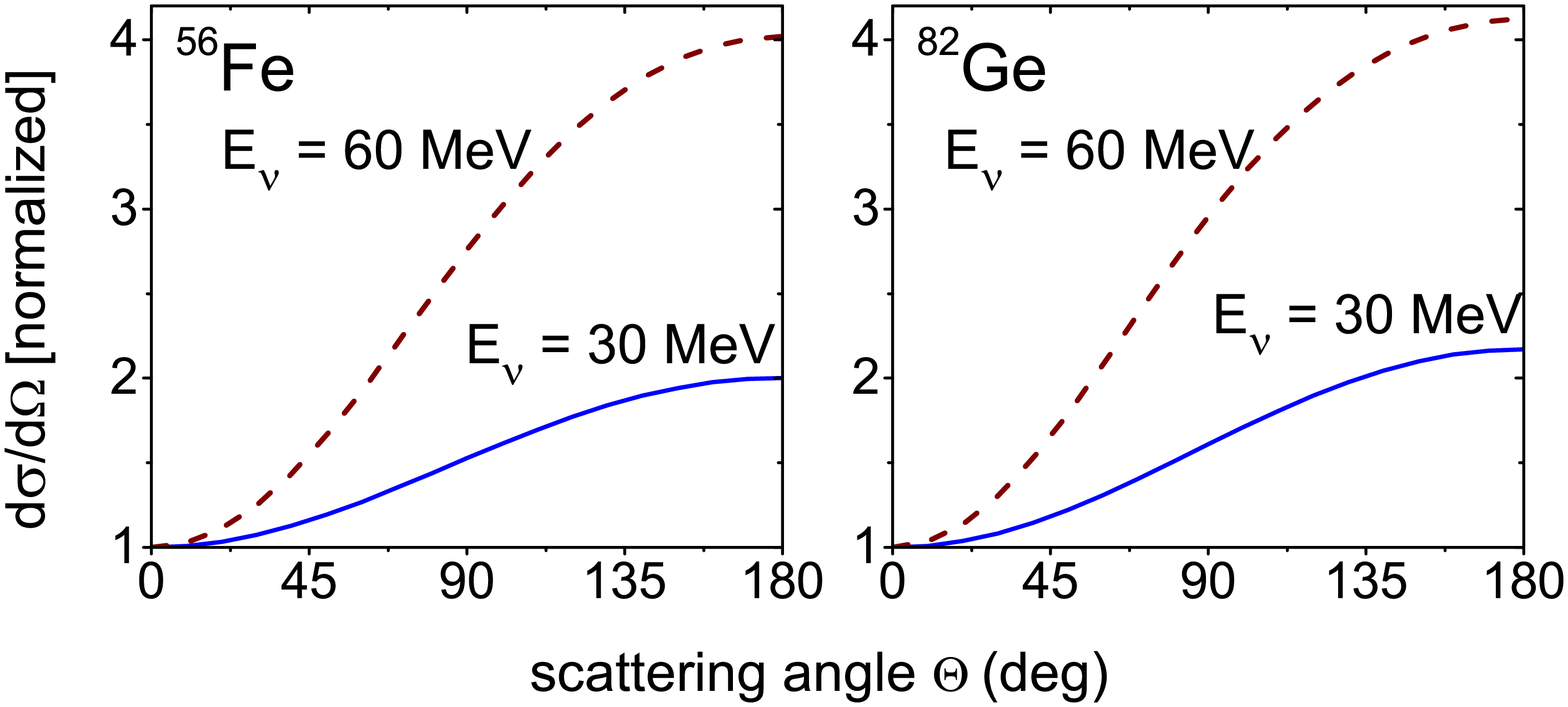}
\caption{(Color online) Normalized differential  cross-sections as a function of the scattering angle. }
\label{angular}
 \end{centering}
\end{figure}

The angular distributions of the scattered neutrinos are shown in
Fig.~\ref{angular} for two incoming neutrino energies,
$E_\nu=30~\mathrm{and}~60\,\mathrm{MeV}$. To make the presentation
more transparent, we normalize the calculated differential  cross
sections to their value at $\Theta=0^\circ$. As shown in the figure,
neutrinos scatter predominately in the backward direction. For
$E_\nu=30\,\mathrm{MeV}$, when $1^+$ transitions dominate, the momentum transfer
is small and the angular
dependence of the differential cross-section essentially corresponds to
$d\sigma/d\Omega\sim (1 + \sin^2(\Theta/2))$~\cite{Donnelly_PRep50}.  The
small deviation for $^{82}$Ge is due to a
non-negligible contribution from the forbidden multipoles (see Fig.~\ref{Mult_contributions}). For
$E_\nu=60\,\mathrm{MeV}$ neutrinos, owing to the dominant
contribution of forbidden multipoles, the backward-to-forward
asymmetry of the differential cross-sections becomes more
pronounced.

\subsection{Finite temperatures}

\begin{figure}[t!]
 \begin{centering}
\includegraphics[width=\columnwidth, angle=0]{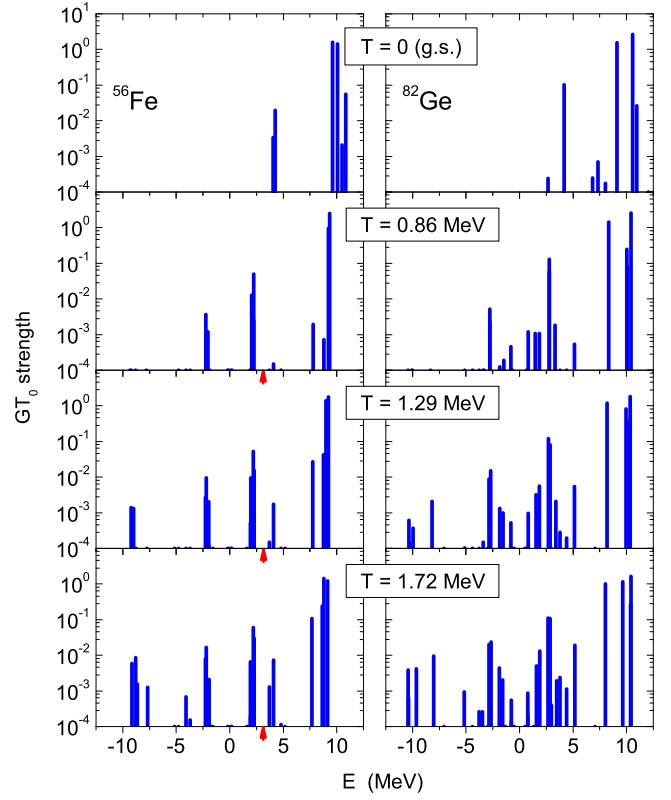}
\caption{(Color online) Temperature evolution of GT$_0$ strength  distributions for
$^{56}$Fe (left panels) and for $^{82}$Ge (right panels) vs.
transition energy. The latter is equivalent to the neutrino energy
transfer. The arrows indicate the zero-temperature threshold
$E_{\mathrm{exp}}({1^+_1}) = 3.12\,\mathrm{MeV}$ for inelastic neutrino scattering off $^{56}$Fe.} \label{GT0_T}
 \end{centering}
\end{figure}

Now we turn our discussion to  thermal effects on the inelastic
neutrino-nucleus scattering. We start by considering the temperature
evolution of the strength distributions for GT$_0$ transitions which dominate
low-energy neutrino scattering. In Fig.~\ref{GT0_T}, we display on a
logarithmic scale the GT$_0$ strength distributions at three different
temperatures relevant in the supernova
context~\cite{Juodagalvis_NPA747}: $T=0.86$~MeV corresponds to the
condition in the core of a presupernova model for a
$15\text{M}_{\odot}$ star; $T=1.29$~MeV and $T=1.72$~MeV relate
approximately  to neutrino trapping and neutrino thermalization
stages, respectively. The transition energy $E$ refers to the
excitation energy of a thermal one-phonon state and is equivalent to
the neutrino energy transfer. To make the thermal effects more visible,
the ground-state GT$_0$  distributions are displayed in
Fig.~\ref{GT0_T} as well.

Since the Brink hypothesis is
not valid within our approach,  in Fig.~\ref{GT0_T} we  observe a redistribution of
the GT$_0$ strength for upward transitions ($E>0$). Namely, at temperatures above the critical one
no extra energy has be paid to break a Cooper pair.  Therefore, by virtue of the vanishing of
pairing correlations, both the GT$_0$ resonance and its low-energy
tail move to lower  energies. Our calculations indicate that, with increasing
temperature up to 1.72~MeV, the resulting resonance
energy shift reaches about 1.5~MeV in $^{56}$Fe and 1.2~MeV in $^{82}$Ge.
It is interesting to note that within the present TQRPA calculations for $^{56}$Fe the
low-lying GT$_0$ strength shifts
below the zero-temperature threshold (i.e., below the experimental energy of the first $1^+$ state).
Furthermore, the thermal smearing of the nuclear Fermi surface makes
low-energy particle-particle and hole-hole transitions possible
which are Pauli-blocked at zero temperature. Such
thermally-unblocked transitions  enhance the low-lying component of
the GT$_0$  distributions and make it more fragmented. Since the
$^{82}$Ge nucleus has a larger single-particle level density near
the Fermi surface, the temperature-induced enhancement and fragmentation of the
low-lying GT$_0$ upward strength is more significant than
in $^{56}$Fe.

\begin{figure}[t!]
 \begin{centering}
\includegraphics[width=\columnwidth, angle=0]{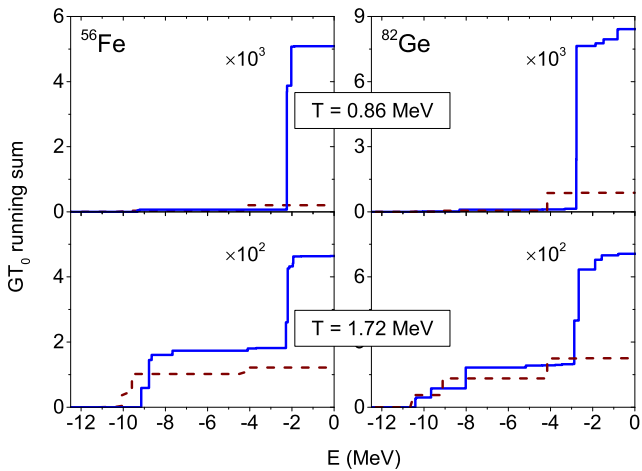}
\caption{(Color online) Comparison of the running sums for GT$_0$ downward strength distributions obtained
using (dashed lines) and without
using (solid lines) the Brink hypothesis. Note,
that the values are scaled by a factor of $10^3$ ( $T=0.86\,\mathrm{MeV}$) and $10^2$ ($T=1.72\,\mathrm{MeV}$).} \label{GT0_running}
 \end{centering}
\end{figure}

Here we would like to stress that the appearance of a sizeable amount of the low-lying transition
strength in nuclei at $T \neq 0$ is predicted in all theoretical
studies of hot nuclei. For example, this was found already in one of
the first papers on the subject Ref.\,\cite{Vautherin_NPA422}, where
the distributions of the electric $E1$ and $E3$ transitions at $T \neq 0$ in
$^{208}$Pb were calculated and in many subsequent studies (see, e.g.,
\cite{Bortignon1998,Santonocita_EPJA30,Niu_PLB681}. The same effect is predicted for
the charge-exchange allowed and first-forbidden transitions as
well \cite{Dzhioev_PRC81,Cooperstein_NPA420}. For the charge-exchange Gamow-Teller
transitions this feature was also
obtained within the shell-model Monte-Carlo
theory~\cite{Radha_PRC56}. Moreover, shell-model Monte-Carlo calculations demonstrate that with
increasing temperature the centroid of the GT$_+$ resonance shifts to lower energies.

Focusing our attention on the negative energy downward transitions  we observe
from Fig.~\ref{GT0_T} that the  corresponding GT$_0$  strength increases with increasing temperature.
This is just a consequence of detailed balance~\eqref{balance}:  the higher the temperature, the more
substantial is the population of nuclear excited states and hence,
the higher is the probability to gain energy from a hot nucleus. Note that the GT$_0$ strength around
$E_\nu\approx - 9\,\mathrm{MeV}$ can be attributed to the
deexcitation of the GT$_0$ resonance. Given the importance of thermal effects on the upward strength distributions, it is worthwhile  to examine how the violation of Brink's hypothesis affects the
downward strength. It is obvious that the shift of the GT$_0$ distributions to lower energies and the
appearance of low-energy transitions
should magnify the strength of downward transitions. This effect is clearly demonstrated in Fig.~\ref{GT0_running} which shows
running sums for  the GT$_0$ downward strength distributions derived by using the Brink hypothesis
or not. The former are obtained from the ground-state ($T=0$) distributions by multiplication
with the Boltzmann factor $\exp(-E/T)$.  Referring to the figure, the considerable increase of the  overall downward strength is mainly caused
by the thermal effects on the low-energy  tail in the GT$_0$ distributions. This is most pronounced at low temperatures ($T=0.86\,\mathrm{MeV}$).
However, at high temperatures ($T=1.72\,\mathrm{MeV}$) the GT$_0$ resonance becomes thermally populated and its shift to lower energies also contributes to the downward strength
increase.

\begin{figure}[t!]
 \begin{centering}
\includegraphics[width=\columnwidth, angle=0]{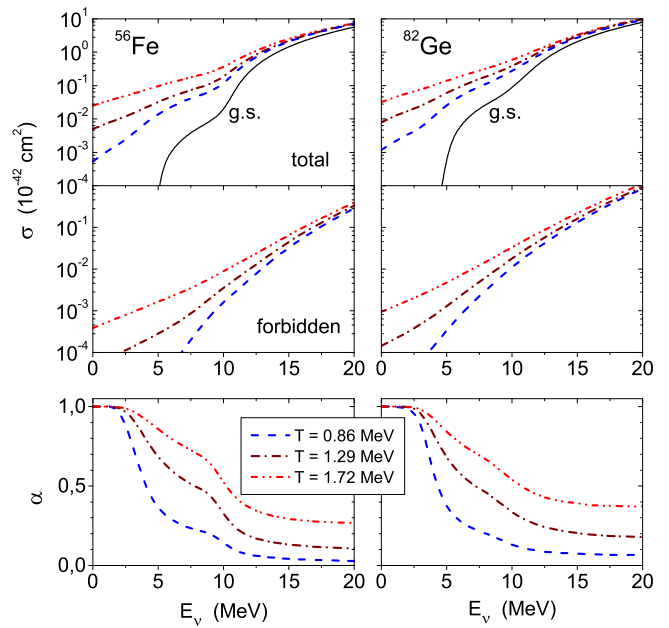}
\caption{(Color online) Upper panels:  Total inelastic  neutrino scattering cross
sections for $^{56}$Fe and $^{82}$Ge at three different temperatures relevant
for core-collapse. For comparison
 the ground-state cross-sections are also shown.
 Middle panels: Contributions of forbidden transitions to the finite-temperature cross-sections.
 Lower panels: Temperature dependence of the fraction of
 down-scattered neutrinos in the thermal enhancement of the cross-section.}
\label{CrSect_T}
 \end{centering}
\end{figure}

The detailed discussion above allows us to understand better the thermal
effects on the inelastic neutrino-nucleus scattering. In the top panels
of Fig.~\ref{CrSect_T}, we compare the ground-state cross-sections
with those calculated at the three core-collapse temperatures. As
follows from our calculations, temperature effects are unimportant
for $E_\nu > 20$~MeV  when neutrinos have sufficiently large energy
to excite the GT$_0$ resonance and collective excitations with other
multipolarities. Note that a downward shift of the GT$_0$ resonance only
marginally affects the cross-sections at such high neutrino energies.  However,
as one can see from the plots, the cross-sections significantly
depend on temperature for low-energy neutrinos. Namely, the reaction
threshold disappears and the cross-sections are enhanced by up to
two orders of magnitude when the temperature rises from
$0.86\,\mathrm{MeV}$ to $1.72\,\mathrm{MeV}$. It is significant that all these features have
pointed out in~\cite{Fuller_APJ376,Juodagalvis_NPA747,Dzhioev_PAN74}
as well.

In Fig.~\ref{CrSect_T}, we also demonstrate the overall
contribution of the forbidden transitions $J^\pi=0^-,\,1^-,\,2^\pm,\,3^\pm$
to the cross-sections. In contrast to hybrid approach calculations~~\cite{Juodagalvis_NPA747}
their contributions are temperature
dependent. However, comparing the upper and middle panels of
Fig.~\ref{CrSect_T}, one concludes that the enhancement of the cross-sections
at finite temperatures is essentially due to thermal effects on the GT$_0$ transition strengths.

At vanishing neutrino energies, $E_\nu\approx 0$, the finite temperature cross-sections
are given by the second, up-scattering, term in Eq.~\eqref{cr_sect_GT0} which accounts for the GT$_0$ downward transitions from thermally excited nuclear states.  As shown in
Fig.~\ref{GT0_T} and pointed out previously, the strength of such transitions increases with
temperature thereby enhancing  the cross-sections. However, in our approach due to the violation of
the Brink hypothesis, the down-scattering part of the cross-section,
$\sigma_\mathrm{down}$, is also temperature dependent and it increases with temperature owing to the
thermally unblocked low-energy GT$_0$ transitions and the
downward shift of the GT$_0$ resonance. This effect is
clearly shown in Ref.~\cite{Dzhioev_PAN74} for $^{54}$Fe.

To analyze relative importance of  the two types of neutrino
scattering processes in the thermal enhancement  of the cross-section,
we introduce the ratio $\alpha$
\begin{equation}
\alpha = \frac{\sigma_\mathrm{up}(T)}{\sigma(T) - \sigma_\mathrm{g.s.}},
\end{equation}
where the difference $\sigma(T) - \sigma_\mathrm{g.s.}$ represents an overall enhancement of the cross-section  due to
thermal effects.
Note that within the hybrid approach $\alpha=1$, because $\sigma(T) =\sigma_\mathrm{g.s.}+\sigma_\mathrm{up}(T)$ in this approach.  We plot
the ratio $\alpha$ in the lower panels of Fig.~\ref{CrSect_T} as a
function of $E_\nu$ for the selected temperatures. As expected, the
ratio~$\alpha\sim 1$ for low-energy neutrinos and then, with increase of $E_\nu$, $\alpha$ gradually decreases indicating a rising contribution  of the up-scattering process to the cross-section thermal enhancement.
It is seen from the plots that for $5\,\mathrm{MeV} < E_\nu < 10\,\mathrm{MeV}$ neutrinos, both from the
up-scattering and down-scattering processes contribute to the noticeable
enhancement of  the cross-sections, although their relative importance depends on temperature:
the higher the temperature the more important is the contribution of the up-scattering process.
Consequently, even for $E_\nu\approx 10\,\mathrm{MeV}$ neutrinos,  when the excitation of the GT$_0$ resonance becomes possible,
the up-scattering component of the cross-section appears to be comparable with the down-scattering one for temperatures $T\geq1.29\,\mathrm{MeV}$.

\begin{figure}[t!]
 \begin{centering}
\includegraphics[width=0.85\columnwidth, angle=0]{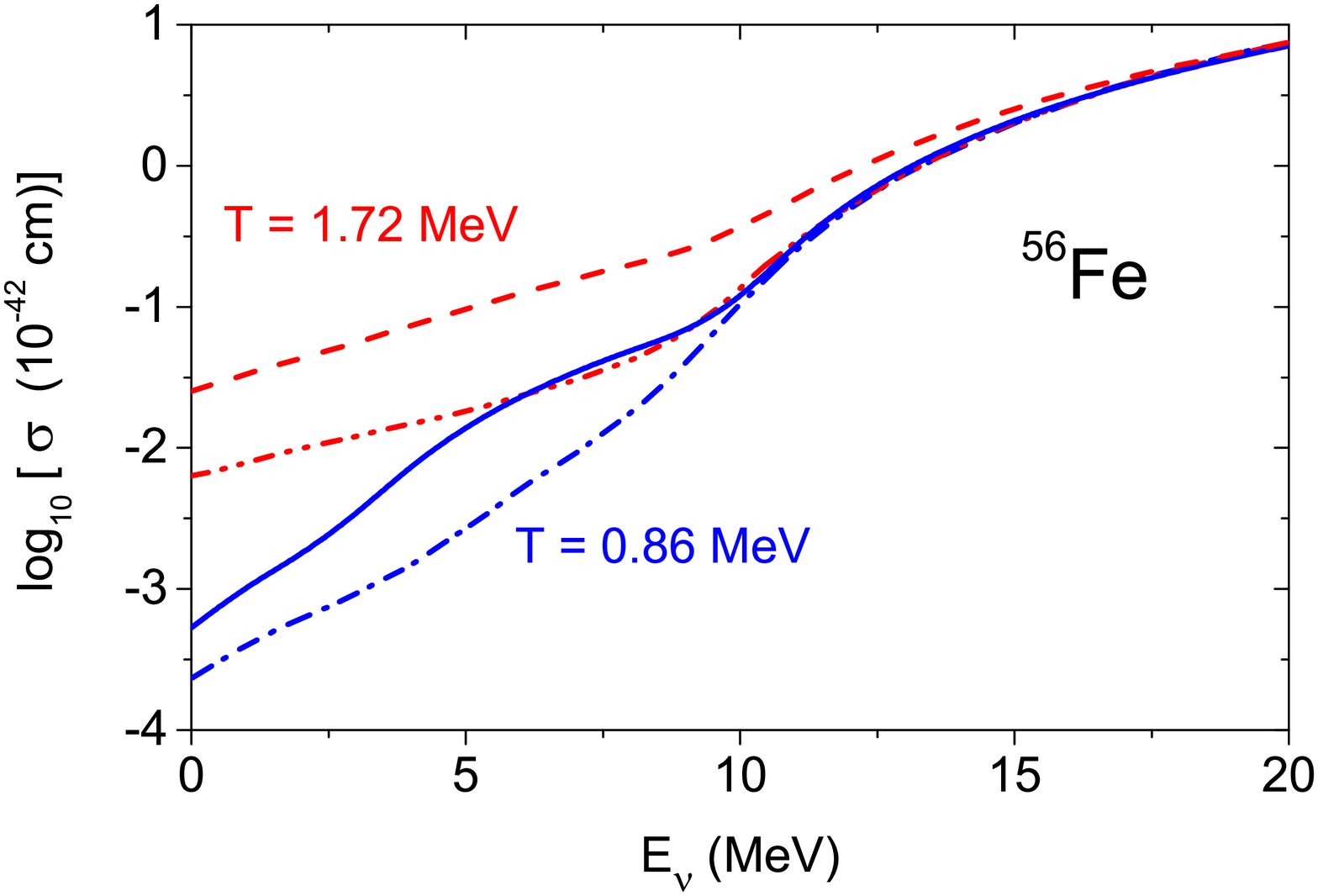}
\caption{(Color online) Comparison of the cross-sections of neutrino inelastic
scattering off the hot nucleus $^{56}$Fe calculated within the
present TQRPA approach and the hybrid approach
(Ref.~\cite{Juodagalvis_NPA747}, Fig.\,11). The solid and dashed
lines show the present results for $T=0.86\,\mathrm{MeV}$ and
$T=1.72\,\mathrm{MeV}$, respectively. The dash-dotted and
dash-double-dotted lines show results from
Ref.~\cite{Juodagalvis_NPA747} for the same $T$ values.}
\label{CrSect_T_comparison}
 \end{centering}
\end{figure}

\begin{figure}[t!]
 \begin{centering}
\includegraphics[width=\columnwidth, angle=0]{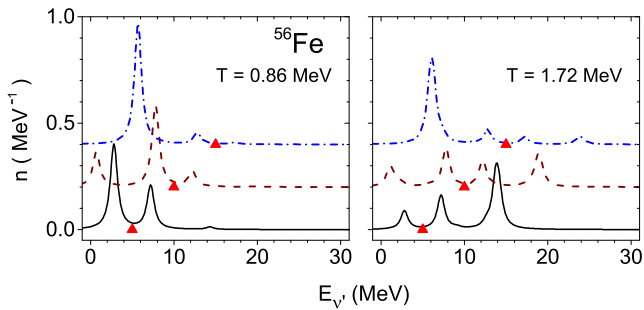}
\caption{(Color online) Normalized spectra of outgoing neutrinos for $^{56}$Fe at $T = 0.86\,\mathrm{MeV}$ and $1.72\,\mathrm{MeV}$
  and three initial neutrino energies: $E_\nu=5\,\mathrm{MeV}$ (solid line),
  $10\,\mathrm{MeV}$ (dashed line, all values shifted by~0.2), and $15\,\mathrm{MeV}$  (dash-dotted line,
  all values shifted by 0.4).  The triangles correspond to the energy of the incoming neutrino. }
\label{spectra}
 \end{centering}
\end{figure}

In Fig.~\ref{CrSect_T_comparison}, we compare our results for
$^{56}$Fe with those obtained within the hybrid approach
\cite{Juodagalvis_NPA747}. The comparison is made for
temperatures $T=0.86~\mathrm{and}~1.72~\mathrm{MeV}$. As one can see, at
$E_\nu<10\,\mathrm{MeV}$ there is noticeable disagreement
between the results of the two approaches: The TQRPA cross-sections are larger by a
factor of 2 to 5  than the hybrid approach ones. To understand the cause of the discrepancy, we calculate the
spectrum of outgoing neutrinos scattered off $^{56}$Fe at the same temperatures as in Fig.~\ref{CrSect_T_comparison} and
compare the results with the hybrid approach calculations (see Fig.\,13 of Ref.~\cite{Juodagalvis_NPA747}).
In Fig.~\ref{spectra},  the spectra are shown
for the same initial neutrino energies as in Ref.~\cite{Juodagalvis_NPA747}: $E_\nu=5,\,10,~\mathrm{and}~15~\mathrm{MeV}$.
Note that for a clearer presentation and for comparison convenience,
the TQRPA spectra are normalized to unity and
folded with the Breit-Wigner function with a
width of $1\,\mathrm{MeV}$.

At low temperatures, the downward transitions from the thermally excited GT$_0$ resonance are strongly suppressed by the Boltzmann factor.
Therefore, for $T=0.86\,\mathrm{MeV}$ and ${E_\nu = 5\,\mathrm{MeV}}$ the spectrum is dominated by neutrinos up- and down-scattered due to the low-energy GT$_0$
transitions.  In Fig.~\ref{spectra}, such transitions correspond to the sizable peaks in the spectrum at around  $E_{\nu'}\sim
(E_\nu\pm 2.5\,\mathrm{MeV})$. The dominance of low-energy up- and down-transitions in the scattering of low-energy neutrinos  off $^{56}$Fe at $T=0.86\,\mathrm{MeV}$ is also
supported by the hybrid approach studies (see the upper-middle panel of Fig.~13 in Ref.~\cite{Juodagalvis_NPA747}). However, the energy and the strength of such transitions calculated with the TQRPA and the hybrid approach are different.
As discussed in detail above, within the TQRPA thermal effects shift the low-lying GT strength in $^{56}$Fe to energies
below the zero-temperature  threshold (Fig.~\ref{GT0_T}) and significantly increase the strength of the corresponding inverse downward transitions (Fig.~\ref{GT0_running}),
thus favoring  neutrino inelastic scattering.
No such effects are expected within the hybrid approach due to the application of the Brink hypothesis. For this reason, the low-energy ($E_\nu < 10\,\mathrm{MeV}$) TQRPA cross-section at $T=0.86\,\mathrm{MeV}$ appears to be larger than the hybrid approach one. With increasing energy of incoming neutrinos, the excitation of the  GT$_0$ resonance comes into play
as evidenced by peaks in the spectra around $E_{\nu'}\sim (E_\nu\ - 9\,\mathrm{MeV})$. Consequently,
the cross-section becomes less sensitive to thermal effects on the low-lying GT$_0$ strength. As a consequence for $E_\nu > 10\,\mathrm{MeV}$ we observe  excellent agreement between the TQRPA and the hybrid approach results.

The situation is slightly different for the higher temperature, $T=1.72\,\mathrm{MeV}$.
Now downward transitions from the thermally excited GT$_0$ resonance  are possible, and
owing to a large phase-space factor they can contribute significantly to the up-scattering of low-energy neutrinos.
In the spectra this contribution corresponds to the peak at around $E_{\nu'}\sim (E_\nu\ + 9\,\mathrm{MeV})$.
As indicated in Fig.~\ref{GT0_running},  at $T=1.72~\mathrm{MeV}$ thermal effects increase the strength of downward transitions from
the  GT$_0$ resonance. Although this increase is not of the same magnitude as for the low-lying GT$_0$ strength, due to a larger phase-space factor,
its contribution to the cross-section enhancement is substantial.
Therefore, one can conclude that at $T=1.72\,\mathrm{MeV}$  the joint action of thermal
changes in both the low-lying GT$_0$ strength and  the GT$_0$
resonance enhances the absolute value of the TQRPA cross-section at
$E_\nu\lesssim 10~\mathrm{MeV}$ in comparison with the hybrid approach.
Like for the $T=0.86\,\mathrm{MeV}$ case, thermal effects become less important for higher neutrino energies,
and both approaches yield very similar results for the cross-sections.

Considering thermal effects on angular distributions of outgoing
neutrinos we find that they are rather unimportant. For up-scattered
neutrinos these distributions are only slightly more backward peaked
than those for down-scattered neutrinos owing to the larger momentum
transfer.

\section{Conclusion}\label{conclusion}

We have studied thermal effects on the inelastic neutrino-nucleus
scattering in the supernova environment. The thermal effects were
treated within the thermal quasiparticle random phase approximation.
The calculations were performed for $^{56}$Fe and $^{82}$Ge.

We find that the TQRPA calculations do not support the Brink
hypothesis and lead to temperature-dependent strength distributions
for both allowed and forbidden transitions involved in the neutrino
inelastic scattering. It is shown that thermal effects shift the
GT$_0$ centroid to lower energies and make low-energy
GT$_0$ transitions possible. As a result, in contrast to hybrid approach
calculations~\cite{Juodagalvis_NPA747}, both the up-scattering and
down-scattering components of the cross-section exhibit a
noticeable temperature dependence at low-energy neutrinos.

Our calculations reveal the same thermal effects as the hybrid approach
based calculations.  Namely, the reaction threshold for inelastic
neutrino-nucleus scattering is removed at finite temperatures and
the cross-section for low-energy neutrinos is significantly
enhanced.

However, the calculated cross-sections for $^{56}$Fe at low neutrino
energies are several times larger than those obtained within the
hybrid approach. We have shown that the discrepancy is due to the violation of
Brink's hypothesis in our approach. This is the main result of the
present study. In addition, it has been demonstrated that the TQRPA
approach can be used to study inelastic neutrino
scattering off massive neutron-rich nuclei at finite temperatures.
Another advantage of our approach is that it incorporates the detailed balance principle, whereas
in the hybrid approach detailed balance is  violated.

There are several directions to improve our approach. At present,
its predictive power is limited by the phenomenological Hamiltonian
with schematic separable residual interactions. It would
therefore be desirable  to combine our TFD-based TQRPA method with
self-consistent QRPA calculations based on more
realistic effective interactions. For neutrino scattering and
neutrino absorption reactions at zero temperature,  such
calculations were performed recently within the relativistic nuclear
density functional theory~\cite{Djapo_PRevC86,Paar_PRevC87}. This
improvement would also allow to take into account the effects of
nuclear deformation. For supernova electron-capture rates in
$pf$-shell nuclei, deformation was recently included in
self-consistent QRPA calculations with the Skyrme
interaction~\cite{Sarriguren_PRevC87}. The other direction of
improvement is the inclusion of correlations beyond  the TQRPA by
taking into account the coupling of thermal phonons with more
complex (e.g. two-phonon) configurations. At zero temperature this
problem was considered within the QPM~\cite{Soloviev1992} by exploiting
separable schematic effective interactions. However, with the
separable approximation for the Skyrme interaction~\cite{Giai_PRevC57}
one could consider phonon coupling at finite temperatures within a
self-consistent theory.

\begin{acknowledgments}

The collaboration with Prof.~T.~S.~Kosmas and Dr.~V.~Tsakstara
from the University of Ioannina is acknowledged.
We are also greatly indebted to Prof. G.~Mart\'{\i}nez-Pinedo
for helpful discussions and important comments on this paper.
This work was supported in part by the Heisenberg-Landau Program,  the
CNRS-RFBR grant 11-091054, and the Deutsche Forschungsgemeinschaft grant No. SFB 634.

\end{acknowledgments}



%

\end{document}